\documentclass[12pt,epsf]{article}
\newcommand{\bfpar}{\mathop{\partial}^\leftrightarrow}

\newcommand{\KK}{{\cal K}}
\newcommand{\LL}{{\cal L}}
\newcommand{\OO}{{\cal O}}

\newcommand{\TT}{{\cal T}}

\newcommand{\UU}{{\cal U}}

\newcommand{\wh}{\widehat}

\newcommand{\be}{\begin{equation}}
\newcommand{\ee}{\end{equation}}
\newcommand{\ben}{\begin{eqnarray}\displaystyle}
\newcommand{\een}{\end{eqnarray}}
\newcommand{\refb}[1]{(\ref{#1})}
\newcommand{\p}{\partial}
\newcommand{\tT}{\widetilde T}
\newcommand{\tF}{\widetilde F}
\newcommand{\tG}{\widetilde G}
\newcommand{\hmu}{{\hat\mu}}
\newcommand{\hnu}{{\hat\nu}}
\newcommand{\al}{\alpha}
\newcommand{\ta}{\widetilde\alpha}
\newcommand{\sech}{\hbox{sech}}
\newcommand{\sectiono}[1]{\section{#1}\setcounter{equation}{0}}

\begin{document}

{}~ \hfill\vbox{\hbox{hep-th/0011226}
\hbox{CTP-MIT-3044}
\hbox{UUITP-11/00}}\break

\vskip 1.cm

\centerline{\large \bf  Gauge Fields and Fermions in Tachyon Effective
Field
Theories}
\vspace*{1.5ex}

\vspace*{4.0ex}

\centerline{\large \rm Joseph A. Minahan\footnote{E-mail:
minahan@mit.edu}}
\vspace*{2.5ex}
\centerline{\large \it Department of Theoretical Physics}
\centerline{\large \it Box 803, SE-751 08 Uppsala, Sweden}
\vspace*{3.0ex}
 \centerline{\large \rm and}
\vspace*{3.0ex}
 \centerline{\large \rm Barton Zwiebach\footnote{
E-mail: zwiebach@mitlns.mit.edu}}

\vspace*{2.5ex}

\centerline{\large \it Center for Theoretical Physics}
\centerline{\large \it Massachusetts Institute of Technology}
\centerline{\large \it  Cambridge, MA 02139, USA}

\vspace*{4.5ex}
\medskip
\centerline {\bf Abstract}

\bigskip
In this paper we  incorporate gauge fields into the tachyon field theory
models for unstable  D-branes
in bosonic and in Type II string theories.
The chosen couplings yield massless gauge fields
and an infinite set of equally
spaced massive gauge fields on
codimension one branes.  A lack of a continuum spectrum 
is taken as evidence that the stable tachyon vacuum does
not support conventional gauge excitations.
For the bosonic string model we find two possible 
solvable couplings, one closely related to Born-Infeld
forms and the other allowing a detailed comparison
to the open string modes on bosonic D-branes.
We also show how to include fermions in the type II model.
They localize correctly on stable codimension one branes
resulting in bose-fermi degeneracy at the massless level.
Finally, we establish the solvability of a large class 
of models that include kinetic terms with more than two derivatives.

\vfill \eject
\baselineskip=17pt

\tableofcontents

\sectiono{Introduction and summary}\label{Intro}

While the complete understanding of
Sen's conjectures on tachyon condensation and D-brane annihilation
\cite{senconj} is likely to require the full use of string field
theory, simplified field theory
models of tachyon dynamics are
a useful tool in describing
some aspects of this dynamics in a simpler context.
In the familiar cubic version of open string field theory
\cite{WITTENBSFT}
(and its nonpolynomial extension to open superstring field theory 
\cite{9503099})
the phenomenon of tachyon condensation is really a
condensation of a {\it string field};
the tachyon, along with
an infinite number of scalar modes acquire expectation values
\cite{cubiccalc,superchecks, insight}.
On the other hand, in boundary string field theory
(B-SFT) \cite{9208027,9303143} --  a much less
developed version of string field theory incorporating a certain
degree of background independence, tachyon condensation
involves the tachyon field alone. Indeed, we proposed two
models of tachyon dynamics \cite{0008231,0009246} that
anticipated the exact tachyon potentials that emerge from analysis
of
B-SFT
for both the open bosonic string
and the open superstring respectively
\cite{0009103,0009148,0010108,0009191}. Our models
were derived by taking an $\ell \to \infty$ limit
of tachyon models \cite{0008227} whose spectrum around
soliton solutions are governed by reflectionless
Schroedinger potentials $U_\ell = \ell^2 - \ell (\ell+1)
\hbox{sech}^2 (x)$.\footnote{For a
pedagogical  review on these and other solvable Hamiltonians with
references to the early literature see \cite{9405029}.
Applications of  reflectionless systems to fermions can be found
in \cite{9408120}.}

We included
 gauge fixed dynamics in the open bosonic string
model of ref. \cite{0008231}. These interactions, inspired by those of
cubic string field theory considered in \cite{0002117},
give the expected
results for the gauge field spectrum
on the brane represented by a lump solution.
Nevertheless, recent results in B-SFT including gauge fields
\cite{0010021,0010028,0010218,0011009,0011033} 
indicate that  simpler gauge invariant
interactions
would also represent the physics
quite accurately. One of the objectives of the present paper is
to introduce into both the bosonic string and the superstring
tachyon models gauge invariant interactions that preserve solvability
and lead to localized gauge invariant dynamics on the world volume
of solitons representing lower dimensional branes. Again, much
of our insight is based on the behavior of finite $\ell$
models.
Happily, our results are consistent with the general
(Born-Infeld type)
forms  discussed in
\cite{0010021,0010028,0010218,0011009,0011033}.

\medskip
We will show that the gauge interactions in the superstring model
lead
to a discrete spectrum for the gauge fluctuations on the
tachyon background representing
the codimension
one brane. In particular we find no continuum sector. The masses for the
fluctuations match the masses found for the tachyon fluctuations.
The coupling of the 
gauge fields 
is of the form $\exp (-T^2/2) F_{\mu\nu}^2$.
Since the tachyon vacuum is at $T=\pm \infty$
one may ask, as usual, whether
the gauge field has dynamics on this vacuum. While the factor
multiplying
this gauge field is going to zero, suggesting 
that the gauge field becomes nondynamical and acts
as a Lagrange multiplier that sets 
the associated
currents to zero \cite{9909062}, this vanishing prefactor could
alternatively be taken as a sign of strong coupling dynamics
\cite{9901159,0002223} in which case the fate of the gauge
field is less clear.
We think, however, that the proof of gauge field localization on the 
codimension one brane, as evidenced by the 
discrete spectrum of gauge field
fluctuations without a continuum, implies
that the tachyonic vacuum does not support conventional local gauge
field excitations.
This is because the asymptotic regions of the  codimension one
brane live on the tachyon vacuum, and particle-like gauge
fluctuations around
this vacuum would have manifested themselves as a continuum spectrum
of fluctuations of the gauge field around the lump solution.

\medskip

For the bosonic case, we discuss two plausible  gauge interaction terms.
In the first one,
arising from a Born-Infeld
action, we find a discrete spectrum, where the spacing
between levels
is twice the spacing for the tachyon fluctuations.
For the second one the tachyon
prefactor in front of the gauge
kinetic term is the same as that
for the tachyon kinetic term.
Here we find that the level spacing  matches
those of the tachyon fluctuations. This second form of the coupling does not
appear to have a solvable finite $\ell$ counterpart. Nevertheless, the
correctness of the mass level spacing leads us to believe that
this is the coupling chosen in string theory.  The bosonic 
Born-Infeld action was considered in the recent papers of
Cornalba \cite{0010021}, Okuyama \cite{0010028}, Andreev \cite{0010218} and
Tseytlin \cite{0011033}.  In particular, Andreev \cite{0010218} discussed the
possibility of ambiguities in such actions. The second form was actually
advocated by Gerasimov and Shatashvili \cite{0011009}, as a natural
choice of coordinates in the space of boundary interactions.
While it seems likely that the two actions considered here
would be related by a field redefinition in the context of an extended full
theory including all higher derivatives, the truncations they 
represent do give different spectra. 

\medskip
Indeed, we use the spectrum associated to this second coupling
to demonstrate
how the fluctuation modes for the tachyon and gauge fields compare with
open string states on bosonic D branes. In a complete
 string field theory model of tachyon
condensation  it is not
enough to show that there is a discrete spectrum about the lump
solutions. One must also
account for the huge degeneracies that occur at the higher
mass levels.
In a theory with just a tachyon, there are no degenerate states
at higher
mass levels.  If other fields are introduced, however,  they
too have
fluctuations about the lump and  modes can be degenerate.
Moreover,  an
interesting and highly suggestive pattern emerges. 
The massless gauge field in the bulk localizes to a massless
gauge field on the codimension one brane, with no extra massless scalar, as
opposed to the case of  Kaluza-Klein reduction, where
a massless scalar accompanies the lower dimensional gauge field. 
String theory, however, still has this massless scalar, and
in our framework it arises as a 
fluctuation mode of the tachyon field, the mode representing the
translation of the brane.
Such a pattern appears to repeat at  higher  mass levels and
we believe it is generic: modes 
removed through localization reappear as fluctuation
modes for lower mass fields. 
In  light of these remarks, we now see that the 
equally spaced  mass levels  of the original
tachyon models are more than just suggestive patterns.
These patterns are necessary in order for the models to 
reproduce the full
open string spectrum on a D-brane when 
additional fields are included.
The result, with proper account of higher derivatives, is 
presumably  B-SFT with all fields. 

\medskip
We also explore the coupling of fermions to the
tachyon model of the superstring. Thus we include, along
with the tachyon and the gauge field, the Majorana spinor
present in the (unstable) D9 brane of IIA. We propose
certain couplings of the tachyon field to the fermions
that result in rather desirable properties. Namely, half
of the fermionic degrees of freedom localize on the
kink solution representing the D8,
 giving rise to a massless
spectrum consistent with supersymmetry. The massive fermion
spectrum is calculable and the mass squared spacing is the
expected one. No continuum spectrum arises and the
fermion becomes infinitely massive at the stable tachyonic
vacuum.

\medskip
Finally, we explore  how higher derivative terms
affect the solvability of the models. Focusing on the model
relevant to the superstring we replace the derivative
term $\partial T \partial T$ by a general function
$f(\partial T \partial T)$ where $f'(0) \not=0$, and show
that solvability is preserved while the precise values
for the mass spacings and ratios of tensions between
branes are altered. In particular, judicious choices of
the form of $f$ allow us to obtain (solvable) 
Born-Infeld  type actions.
In particular we study a Born-Infeld type action proposed
by Garousi \cite{0003122} and Bergshoeff {\it et.al.} \cite{0003221} 
where the tachyon kinetic term is insider the Born-Infeld square
root.  Such an action was
argued to be  consistent with  open string scattering amplitudes and
$T$ duality.  We will find that this action has a kink solution
with vanishing width.  In addition, we find that the fluctuations for the
tachyon field have the correct spacing between mass levels, although the
spacings for gauge fluctuations is half the expected amount.

We will also see that higher derivative terms can lead to regular solutions
for higher codimension branes.
For example, with higher derivative terms there can be even
codimension branes which are tachyon solitons
of the world volume theory
of several D9 brane-antibrane pairs of type IIB. Likewise, there can
be odd codimension branes which are solitons of the world-volume
theory of several unstable D9 branes of type IIA.
In agreement with the observations
of \cite{0010108} such solutions do not appear to require
expectation values for the gauge fields. The kinetic form
of the tachyon field allows
finite energy configurations
even though the asymptotic values of the tachyon wind over
the sphere at infinity.

We find it striking just how simple it is for the tachyon to induce
complete localization for the
gauge fields and fermions on the brane. From the viewpoint
of solitons, we see that the complete localization of the
tachyon relies on the absence of tachyon dynamics at the
stable vacuum, something directly guaranteed by the form
of the tachyon potential. This implies that
the spectrum around the configuration representing the
brane cannot have a continuum sector arising from the tachyon field.
This results in
a soliton spectrum of stringy type. Indeed, with the fluctuation
spectrum
governed by a Schroedinger potential with no continuous spectrum
one must necessarily have an infinite
number of energy levels. The complete localization of
gauge fields and fermions follows by couplings to the
localized tachyon. Again, these fields give rise to
an infinite number of fields on the soliton. We find it
quite remarkable that all this is exactly modeled with
simple interactions of a scalar field,
a gauge field and a fermion.

\medskip
This paper is organized as follows.
In section 2 we show how to include gauge fields through gauge
invariant interactions for general
classes of   models with stable kinks, which include the
two derivative truncation
of the unstable D9 brane in type IIA.
In section 3 we do a similar analysis for  models with unstable
lumps, which include
the two derivative truncation for the tachyon effective field theory in
bosonic string theory.
 In section 4 we compare the results for the bosonic gauge fluctuations
to that
of the bosonic string.  Here, we are able to match  the fluctuation
modes
to particular open string oscillator states.
In section 5 we discuss the inclusion
of fermion fields in the models with kinks.  Finally
in section 6 we consider the
extension of the tachyon superstring model where we include
certain kinds of higher derivative terms (this class includes
the modified Born-Infeld action).
We then use such actions to obtain
nontrivial solutions representing branes of codimension higher than one.
Concluding remarks are offered in section 7.

\sectiono{Coupling gauge fields to the superstring tachyon model}

In Ref.~\cite{0008231}, gauge fixed interactions were added to
the bosonic tachyon effective field theory corresponding to the
$\ell=\infty$
model, and to the finite $\ell$ models. Such interactions
were  inspired by similar terms appearing in gauge fixed cubic
string field theory, and had been considered in \cite{0002117}.
The interactions were tailored to result in gauge fields localized on
the
tachyon
solutions representing lower dimensional  branes.
The spectrum of massive gauge excitations could be solved exactly
and were shown to have
 equally spaced levels, with the correct spacing.
Nevertheless, it is now clear that in boundary string field theory
one should naturally be led to a gauge invariant description of
the interactions. Indeed, the forms of such interactions have
recently been discussed by several authors
\cite{0010021,0010028,0010218}.

In light of this, we look, for  gauge invariant interactions  of a gauge
field
with the tachyon field that lead to integer spacings for the gauge
fluctuations,
a residual gauge invariance on the solitons or lumps
and absence of conventional degrees of freedom for the gauge
field in the vacuum.  There are remarkably simple actions that
posess these features for both the bosonic model and the superstring
model. In this section we will discuss the case of the superstring model
while in section 3 will discuss the case of the
 the bosonic string model.  In both cases it will be
instructive to consider the finite $\ell$ models whose limit
$\ell\to \infty$ led to the string models.

\subsection{The superstring model revisited}

Our construction will be quite general, allowing us to specialize to
various models, such as the finite $\ell$ models discussed in
\cite{0008231}, or to the harmonic oscillator model which is reached by
taking the $\ell\to\infty$ limit and which leads directly to the
two derivative truncation of boundary string field theory.
 As shown in
\cite{0009246}, these models can be constructed in generality
by reconstructing the field theory from the profile of the kink
solution. Indeed,
given  a kink profile
\be
\overline \phi (x) = {\cal K}(x) ,
\ee
where  ${\cal K}(x)$, for kink\footnote{In \cite{0009246} we used the
variable
${\cal P}$, for profile, rather than ${\cal K}$. We have changed
notation to
be able to call the lump profiles ${\cal L}(x)$.}, is
either a monotonically increasing or a monotonically decreasing function
of the
coordinate $x
\in [-\infty,
\infty]$, the desired field theory is given by:
\be
\label{finmod}
S = - \int dt d^{p+1}x \Bigl({\cal K}' ( T)\Bigr)^2 \,
\Bigl(\,  (\partial T)^2  + 1 \,\Bigr)\,,
\ee
where the prime denotes partial derivative with respect to the argument,
the scalar field
$T$ is related to $\phi$ as
\be
\label{redkink}
\phi = {\cal K}(T) \,,
\ee
and the kink is just
\be
\label{fkink}
\overline T (x) = x\,.
\ee
The action \refb{finmod}, assuming $\KK''(0)=0$, has a vacuum at $T=0$
where there is a tachyon of mass squared
\be
\label{thetachmass}
M^2_T = {\KK'''(0)\over \KK'(0)}\,,
\ee
In addition, one easily confirms that
\be
{\sqrt{2}\over 2\pi} {\TT_{kink}\over \TT} = {\sqrt 2\over \pi}
\int_{-\infty}^\infty \,
dx \Bigl( \,{\KK'(x)\over \KK'(0) } \,\Bigr)^2\,,
\ee
where $\TT$ denotes the tension of the spacefilling brane represented by
the $T=0$ vacuum of \refb{finmod} and $\TT_{kink}$ is the tension of
the (codimension one) kink.  In string theory the above ratio is one.

For example the
finite $\ell$
models can be obtained by taking
\be
\label{kl}
{\cal K}'_\ell (x) = \sqrt{\TT}\hbox{sech}^{\ell}(x) \,.
\ee
The spectrum of fluctuations around the kink solutions of these
models are governed by a Schroedinger equation with potential
\cite{0008231}
\be
\label{ul}
U_\ell (x)=\ell^2-\ell(\ell+1)\hbox{sech}^2(x) \,.
\ee

We now consider the fluctuation problem around the kink solution
$\overline T(x) = x$.  For this purpose,
we let $T\to x + \tT$ in \refb{finmod}, where $\tT$ represents the
fluctuation
field.
It is then useful to redefine the fluctuation field
as
\be\label{fredkink}
\KK'(x) \tT = \wh T\,.
\ee
Up to quadratic fluctuations  and after dropping
constant terms, the action is
\be
\label{sfuct}
S_{quad} = - \int dt d^{p}y dx \,
\Bigl\{  \,\partial_\hmu \wh T\,
\partial^\hmu \wh T
+  \, \wh T \Bigl( - {\partial^2 \over \partial x^2 }
+ {\KK'''(x)\over \KK'(x)}
\Bigr) \, \wh T \Bigr\}\,.
\ee
The $p+2$ indices
$\mu$ have been split into
 $p+1$ indices $\hat\mu$
along the brane worldvolume $(t, y^i)$, and the index $x$ along the
coordinate
transverse to the brane.
{}From the above equation it follows
that
the Schroedinger problem for the tachyon fluctuations is simply
\be\label{tachschr}
- {d^2\over dx^2 } \wh  T +  {\KK'''(x)\over \KK'(x)} \,
\wh  T = M^2 \wh T\,.
\ee
The reader can verify that with $\KK_\ell$ given in \refb{kl}
the potential term in the above equation equals
$U_\ell$ (\refb{ul}).
It is also clear from the above equation  that $\wh T_0(x) =
\KK'(x)$
is a solution with $M^2=0$.
This is the translation mode of the kink in the original
field variable.
Indeed, from \refb{redkink}, \refb{fkink} and \refb{fredkink}
we have
$\phi = \KK (x + \epsilon \wh T_0(x)/ \KK'(x)) =
\KK(x) + \epsilon \KK'(x) + \OO (\epsilon^2)$.
Since $\KK(x)$ is monotonic, $\KK'(x)$ does not have zeroes.
Therefore, $\KK'(x)$ is the ground state wavefunction for the
Schroedinger
equation in \refb{tachschr} and thus there are no tachyonic
fluctuations.

\subsection{Coupling to gauge fields}

In string theory, the gauge fields appear through the Born-Infeld action
\be\label{gaugebos}
S_{BI}=-\TT\int dt d^{p+1}x\,\, V(T)\sqrt{-\det(\eta_{\mu\nu}
+ F_{\mu\nu})}\,,
\ee
where $V(T)$ is the tachyon potential. Here, we are only interested in
finding the gauge fluctuations, so it suffices to make the approximation
\be\label{sqrap}
\sqrt{-\det(\eta_{\mu\nu}
+ F_{\mu\nu})}=1+\frac{1}{4}F_{\mu\nu} F^{\mu\nu}+... \,.
\ee
Therefore, the action we use for the gauge fields and tachyons is
\be
\label{finmodg}
S = - \int dt d^{p+1}x \Bigl({\cal K}' ( T)\Bigr)^2 \,
\Bigl(\,  (\partial T)^2  + 1 \, + {1\over 4}
F^{\mu\nu}
F_{\mu\nu} \,\Bigr)\,.
\ee

In order to see that \refb{finmodg} is a reasonable coupling we
study the spectrum of fluctuations of the gauge field
on the background of the kink solution. We choose the axial
gauge condition
\be
\label{axial}
A_x = 0 \,, \quad  \to  F_{x\hat\mu} = \partial_x A_{\hat\mu}\,.
\ee
We also note for later reference the equation of motion obtained
by varying $A_x$, which in the chosen gauge simplifies to
\be
\label{misseqn}
\partial_{\hat\mu} \Bigl( (\KK'(T))^2 F^{x \hat\mu}\Bigr) = 0 \quad
\to \quad  \partial_x (\partial_{\hat\mu} A^{\hat \mu}) =0\,,
\ee
where in the last step we have considered the reduction to
linearized equations for gauge
fluctuations on the background of the tachyon kink (the tachyon
expectation value is independent of the brane coordinates).
The term in the action \refb{finmodg} relevant to the gauge field
fluctuations is
\be
\label{hyt}
S(\overline T, A)  = -  \int dt d^{p}y dx\,\Bigl({\cal K}'
(x)\Bigr)^2 \,
\Bigl(\,
{1\over 4} F_{\hat\mu\hat\nu}F^{\hat\mu\hat\nu}
+ {1\over 2}\partial_x A_{\hat\mu}\partial_x
A^{\hat \mu}\,\Bigr) \,.
\ee
The analysis is helped by the field redefinition
\be
\label{gfredef}
B_\hmu = \KK'(x) A_\hmu\, ,\quad \tF_{\hmu\hnu}=
\p_\hmu B_\hnu-\p_\hnu B_\hmu\,,
\ee
which allows us to turn \refb{hyt} into
\be
\label{hytt}
S(\overline T, B(A))  = -  \int dt d^{p}y dx\,
\Bigl(\,
{1\over 4} \tF_{\hat\mu\hat\nu}\tF^{\hat\mu\hat\nu}
+ {1\over 2} \, B_\hmu \Bigl( - {\partial^2 \over \partial x^2 }
+ {\KK'''(x)\over \KK'(x)}
\Bigr) \, B^\hmu  \, \Bigr) \,.
\ee
Comparing this last equation with \refb{sfuct} we see that
the Schroedinger problem governing the spectrum of the gauge
field fluctuations on the kink is the {\it same}
 as that governing
the tachyon fluctuations. Just as before the ground state
is associated to the wavefunction $\KK'(x)$.
Note that the constant in front in \refb{hytt} plays no special role
in this or the following analysis.

It is straightforward to understand the gauge invariance
of the massless gauge mode living on the lump solution.
While we have imposed the axial gauge $A_x=0$,
there is still a residual
gauge invariance under
\be
A_\hmu\to A_\hmu +\p_\hmu \varepsilon\,, \quad \hbox{with}
\quad \p_x\varepsilon=0\,.
\ee
This means that the field $B_\hmu$ introduced
in \refb{gfredef} transforms as
\be
B_\hmu\to B_\hmu+\KK'(x)\,\p_\hmu\varepsilon\,.
\ee
The massless gauge field $B_\hmu (y)$ living on the brane
arises as $B_\hmu (y, x) = B_\hmu (y) \KK'(x)$. It thus follows
from the last equation that
\be\label{Bgauge}
B_\hmu(y) \to B_\hmu (y) +\p_\hmu\varepsilon\,, \quad
\varepsilon=\varepsilon (y) \,,
\ee
which is the standard gauge transformation of a gauge field.
Since for this massless mode $A_\hmu = B_\hmu (y)$ (see \refb{gfredef})
the subsidiary field equation in \refb{misseqn} is identically
satisfied.

We can also confirm that the excited wave
functions arising from the Schroedinger
problem of \refb{hytt} correspond to massive gauge
fields on the brane.  First note that the action
for a mode
\be\label{Bpsi}
B^{\hmu (n)}(y) \psi^{(n)}(x)\,,
\ee
where the wave function $\psi^{(n)}$ has
eigenvalue $M_n^2$, is of the form
\be\label{BFaction}
\int dtd^py\left(\frac{1}{4} (F_{\hmu\hnu}^{(n)})^2
+ \frac{1}{2} M_n^2 (B_\hmu^{(n)})^2\right).
\ee
In addition, the constraint \refb{misseqn},
using \refb{gfredef} and \refb{Bpsi}, reduces to
\be\label{Bdiv}
\partial_x \Bigl( {\psi^{(n)}(x) \over  \KK'(x) }
\Bigr)  \partial_\hmu  B^{\hmu (n)} (y) =0
\, \quad\to\quad  \partial_\hmu  B^{\hmu (n)} (y) =0\,.
\ee
The action \refb{BFaction}, supplemented by this condition
describes a massive gauge field.

\medskip
Let us now turn to the example of the D9 brane.
In \cite{0009246,0010108} it
was argued that the lagrangian for the tachyon field for
up to two derivatives is
\be\label{fsaction}
S=-{\cal T}\int dtd^9x \exp(-T^2/2)\left[\p_\mu T\p^\mu T +
1\right].
\ee
This action can be obtained from \refb{finmod} by setting
\be\label{kinkharm}
\KK'(T)=\sqrt{\TT}\exp(-T^2/4)=
\lim_{\ell\to\infty}\KK_\ell(T/\sqrt{2\ell}).
\ee
Since by construction, $\KK'(x)$ is a ground state wave function,
we see that the corresponding potential
is
$U(x)=x^2/4$.
Given the potential term in \refb{fsaction}, the Born-Infeld action is
\be\label{fsgauge}
S_{BI}=-{\cal T}\int dtd^9x
\exp(-T^2/2)\sqrt{-\det(\eta_{\mu\nu}+
F_{\mu\nu})}.
\ee
The integrand
 can be expanded as in \refb{sqrap} and so the fluctuation spectrum
is found using the preceding arguments.
Therefore, using \refb{kinkharm} and \refb{hytt},
 the action for the gauge fluctuations
about
the kink solution is
\be\label{gaugebos2a}
-{\cal T}\int dt d^{8}ydx
\left(\frac{1}{4}\tF_{\mu\nu}\tF^{\mu\nu}+\frac{1}{2} B_\mu\left(-
\frac{\p^2}{\p x^2} +
\frac{1}{4}x^2-\frac{1}{2}\right) B^\mu\right).
\ee
Hence, the fluctuation spectrum for the gauge fields is
described by a harmonic oscillator,  with the
lowest mode corresponding to a massless state.  The spacing between the
modes is unity, which using \refb{thetachmass} is twice the mass
squared of the tachyon.
As in \refb{Bgauge} there is  a residual
gauge
invariance for the massless vector mode on the stable 8-brane. 

The
absence of a continuum sector of gauge field fluctuations is 
manifest, and we take this as evidence that the tachyonic
vacuum does not support conventional gauge field excitations.
Indeed, in the case of scalar field excitations the logic connecting
complete localization on the brane to lack of excitations on the
stable vacuum is quite direct. The Schroedinger potential for
fluctuations is $U(x) = V''(\overline \phi (x))=
M_{eff}^2(\overline\phi(x))$,  where $V$ is the scalar field potential and
$M_{eff}^2$ denotes the effective mass of the scalar field (computed by
cancelling the linear term, if nonvanishing, with an external source). We then
have that
$U(x\to \pm \infty) = M_T^2$ where
$M^2_T$ denotes the scalar field mass at the stable vacuum. Thus 
a continuum arises in the Schroedinger problem for energies greater than
$M_T^2$. If $M_T^2\to \infty$ there is no continuum.

\sectiono{Coupling gauge fields to the bosonic string tachyon model}

In this section we discuss the general construction for gauge fields
coupled
to the bosonic string model.  We will give two types of interaction
terms.
The first such term
 leads to a solvable spectrum for all finite $\ell$ models.
This type of interaction is derivable from the Born-Infeld action.
The second such term
only has a solvable spectrum for the $\ell=\infty$ model.
However, its spectrum is what one would
expect from open string modes on D branes in bosonic string theory.
This leads us to believe that this is the choice taken by
B-SFT.

\subsection{The bosonic string model revisited}

We begin by describing a general construction of a scalar
field theory given the profile of a lump solution.
We let
\be
\label{lb}
\overline \phi (x) = \LL (x^2)
\ee
define the lump profile.
 By lump, we mean a field configuration where the derivative of the
profile
with respect to the coordinate $x$ vanishes at one point.
The argument of $\LL$
 manifestly incorporates an expected reflection
symmetry. The equation of motion of the lump relates the
potential $V$ in the action to the spatial derivative
of the profile as follows
\be\label{eqofmot}
V (\overline \phi (x)) = {1\over 2} [ \,\overline\phi ' (x) \,]^2
= 2 x^2 [\, \LL'(x^2)\, ]^2\,.
\ee
{}From \refb{eqofmot} it is clear that $x\LL'(x^2)$ must have one
and only one zero.

We now introduce a new field $\phi = \LL (T)$, which in
view of \refb{lb}  implies
\be
\overline T(x) = x^2.
\ee
 This in turn
gives $V(\overline \phi) = 2 \overline T [\, \LL'(\overline T)\,]^2$.
Thus, the action
$S = - \int ({1\over 2} (\partial \phi)^2+ V(\phi))$
becomes
\be\label{bostachaction}
S = - \int dt d^{p+1}x ( \LL'(T))^2 \,\,\Bigl( \,
{1\over 2} (\partial T)^2  + 2 T \, \Bigr)\,.
\ee
For the finite $\ell$ models, we have that
\be
\LL_\ell(x^2)=\sqrt{\TT\over 2T_0}\,\,
{1\over \hbox{sech}^{\ell-1}(\sqrt{T_0})}\,\hbox{sech}^{\ell-1}(x),
\ee  
where $T_0$ is the value for $T$ at the unstable vacuum, where the potential
is a maximum.  The tension $\TT$ is its vacuum energy.

\medskip
To find the fluctuations, we let $T=x^2+\tau$ and define
the fluctuation field as $\wh T=\LL'(x^2)\tau$.  Up to quadratic order,
the action in
\refb{bostachaction} becomes
\be
S = - \int dt d^{p+1}x\left( \,
{1\over 2} \p_\hmu \wh T\p^\hmu \wh T  + 
\frac{1}{2} \wh T\left(-\frac{\p^2}{\p
x^2}+
 \UU(x)\right)\wh T\, \right),
\ee
where
\be\label{UUdef}
 \UU(x)\equiv \,6\,\frac{\LL''(x^2)}{\LL'(x^2)}+
4x^2\frac{\LL'''(x^2)}{\LL'(x^2)}=\frac{1}{x\LL'(x^2)}
\frac{d^2}{dx^2}\bigl(x\LL'(x^2)\bigr).
\ee
Hence,  we see that
the fluctuation spectrum is determined by the Schroedinger equation
\be
-\frac{d^2}{dx^2}\wh T+\UU(x)\wh T=M^2 \wh T.
\ee
It follows from the last equation in \refb{UUdef} that $x\LL'(x^2)$
is an eigenfunction with
$M^2=0$, and so is the translation mode.  Since $x\LL'(x^2)$ has
one zero, this is  the first excited state and therefore there
exists one tachyonic mode.

We can now apply this general analysis to the specific case of the two
derivative truncation of boundary string field theory.
In boundary string field theory
it was argued that the action for up to two
derivatives for the tachyon field is \cite{0009103,0009148}
\be
\label{origaction}
S = - 4 \,e \TT \int dtd^{p+1}x \Bigl\{  {1\over 2} \partial_\mu \phi
\,\partial^\mu
\phi - {1\over 4} \phi^2 \ln \phi^2 \Bigr\}\,.
\ee
This model can  be obtained by taking the limit $\ell\to\infty$ in
the finite $\ell$ models \cite{0008231}.
The field redefinition $\phi= \exp [ -{1\over 4} T ]$ casts this
action
in
the form:\footnote{We choose to work with the form of the 2-derivative
tachyon action
for which the tachyon mass is the correct one.}
\be\label{bsaction}
S_T=-\frac{e{\cal T}}{4}\int dtd^{p+1}x
\exp(-T/2)\left[\, {1\over 2} \p_\mu T\p^\mu T +
2T\right].
\ee
This action has the form in \refb{bostachaction}, with
\be\label{lumpharm}
\LL'(T)=\frac{1}{2}\sqrt{{\cal T}e}\exp(-T/4),
\ee
 and hence has a codimension lump
solution with $T=x^2$.  From \refb{lumpharm} and \refb{UUdef} we see
that
\be
\UU_\infty(x)=
\frac{1}{4}x^2-\frac{3}{2},
\ee
hence, the fluctuation spectrum has a tachyon with  $m^2=-1$
and higher
mass states with integer spacing, $\Delta m^2=1$.
In \cite{0008231} it was shown that the action in \refb{bsaction}
 also has codimension
$d$ lump solutions with $T=\sum_i^d (x_i^2/2)-2(d-1)$.
In this case, the
fluctuation spectrum is governed by a $d$ dimensional
harmonic oscillator.

\subsection{Coupling to gauge fields}

Now let us include contributions from the gauge fields.
We are again
interested in
the Born-Infeld action in \refb{gaugebos},
but for the analysis of the
fluctuation spectrum it is
sufficient to use
\be\label{bgaction}
S = - \int dt d^{p+1}x ( \LL'(T))^2 \,\,\Bigl( \,
{1\over 2} (\partial T)^2  + 2 T+  2T \,\cdot\frac{1}{4}
F^{\mu\nu}F_{\mu\nu}
 \, \Bigr).
\ee
To find the fluctuation
spectrum about the lump solution, we choose the axial gauge $A_x=0$ and
let
\be\label{BArel}
B_{\hmu}=(2T)^{1/2}\LL'(T)A_{\hmu}\,,
\qquad\qquad
\tF_{\hmu\hnu}=\p_\hmu B_\hnu
-\p_\hnu B_\hmu.
\ee
In terms of these fields, the part of the action containing the gauge
fields
becomes
\be
S(\overline T,B(A))
 = - \int dt d^{p}ydx \,\left( \,
\frac{1}{4}\tF^{\hmu\hnu}\tF_{\hmu\hnu}+\frac{1}{2}
B_\hmu\left(-\frac{\p^2}{\p x^2} +\UU(x)\right)B^\hmu
 \, \right),
\ee
where $\UU(x)$ is defined in \refb{UUdef}.
Hence, the fluctuation spectrum for the $B_\hmu$ fields is governed by
the
same Schroedinger equation as for the tachyon fluctuations.

Since the Schroedinger potentials are the same,
it would seem that the lowest
mode for the gauge field is tachyonic.  But this solution needs to be
discarded, along with all solutions that are even under $x\to -x$.
Upon inspection of \refb{BArel}, we see that $A_\mu\sim 1/x$ if $B_\mu$
is
nonzero at $x=0$, which is the case for the even parity modes.
Therefore, the integrand in \refb{bgaction} diverges
as $1/x^2$, and so the even parity modes should be disregarded.
Hence, the lowest mode corresponds to the first excited state
of the Schroedinger equation and thus is massless.

\medskip
Turning to the specific example of the two derivative truncation of
B-SFT,
the Born-Infeld action  has the form
\be\label{gaugebos2}
-\frac {e{\cal T}}{4}\int dt d^{p+1}x\,\,
2T\exp(-T/2)\sqrt{-\det(\eta_{\mu\nu}+ F_{\mu\nu})}.
\ee
This form of the action has been recently  obtained in
BSFT \cite{0010021,0010028,0010218}.
The overall normalization in \refb{gaugebos}
is chosen so that the gauge kinetic term has the
canonical coefficient at the open string vacuum $T=2$, and so matches
the form in \refb{bgaction} with $\LL'(T)$ defined in
\refb{lumpharm}.
Therefore, the term in \refb{gaugebos} expanded to two derivatives
can be written as
\be\label{gaugebos1}
-\int dt d^{p}ydx
e^{-\overline{T}/2}
\Bigl(\frac{1}{4}\tF_{\hmu\hnu}\tF^{\hmu\hnu}+\frac{1}{2}
B_\hmu\left(-\frac{\p^2}{\p x^2}+
\frac{1}{4}x^2-\frac{3}{2}\right)B^\hmu\Bigr),
\ee
where $\overline{T}$ refers to the tachyon field on the lump.
Hence, the components
of the gauge
field also have a fluctuation spectrum given by a
 harmonic oscillator.

As before, the even parity solutions need to be discarded
 since they  correspond to singular configurations for the
gauge field $A_\hmu$.  Hence the lowest mode is massless and
the spacing between mass squared levels is equal to {\it twice}
the magnitude of the mass
squared term of the original tachyon.

\subsection{An alternative coupling to gauge fields}

The double spacing for the levels seen at the end of the last
subsection seems odd and is perhaps a
consequence of our only considering contributions of up to two
derivatives.  Therefore, let us also consider the gauge coupling
\be\label{hjk}
S' = -\int dt d^{p+1}x (\LL'(T))^2
\, {1\over 4}\, F_{\mu\nu}F^{\mu\nu}\, .
\ee
This does not quite have the Born-Infeld form since it is missing a
factor
of $2T$.
Note that this is the simplest kind of term, the same factor
multiplying the tachyon kinetic term is now multiplying the
conventional gauge kinetic term.

To find the spectrum about the codimension one brane, we set $T=x^2$
and define the new gauge field $B_\hmu$ and associated
field strength to be
\be
B_\hmu=\LL'(x^2)A_\hmu\,, \qquad\tF_{\hmu\hnu}=
\p_\hmu B_\hnu-\p_\hnu B_\hmu\,.
\ee
Subtituting back into \refb{hjk} we find
\be\label{mmn}
 -\int dt d^{p}y dx
\Bigl({1\over 4} \tF_{\hmu\hnu}\tF^{\hmu\hnu}+
\frac{1}{2}\p_x B_\hmu\p_x B^\hmu
+\frac{1}{2}U(x) B_\hmu B^\hmu\Bigr),
\ee
where $U(x)$ is given by
\be\label{Udef}
U(x)=2{\LL''(x^2)\over \LL'(x^2)} + 4x^2 { \LL'''(x^2)\over \LL'(x^2)}
= {1\over \LL'(x^2)} {d^2\over dx^2} \LL'(x^2).
\ee
This is a different Schroedinger potential than in \refb{UUdef} and
does not seem to give a solvable spectrum for the gauge fields in
the finite $\ell$ models.

For the $\ell=\infty$ model, however, the coupling in \refb{hjk} still
gives a solvable spectrum for the gauge fluctuations.
Using the expression for $\LL'(T)$ in  \refb{lumpharm}, the gauge action
is
\be
S' =   -{e\TT\over 16}\int dt d^{p+1}x \exp (-T/2)
F_{\mu\nu}F^{\mu\nu}\, .
\ee
This form of the action has been recently advocated in \cite{0011009}.
Using \refb{Udef}, we  have that
\be
U(x)=\frac{1}{4}x^2-\frac{1}{2}.
\ee
Therefore,
in contrast to the previous case, we find that the lowest mode
for this harmonic oscillator equation is massless.
In fact, there is no need to throw out the even parity modes, since
these
do not correspond to singular configurations for $A_\hmu$.
Thus, the spacing between the gauge modes
is the mass squared of the tachyon.

As in the superstring case,
selecting a gauge imposes a constraint on the remaining fields.
Proceeding
as in \refb{Bdiv}
one finds the constraint on the massive modes
\be
\p_x\p^\hmu(\exp(\tT/2) B^{(n)}_\hmu)=0,
\ee
 where $\tT$ refers to the tachyon field for the effective theory
on the lump. For
the effective field theory on the $p$ brane, this constraint on the
modes can be implemented
through a Stueckelberg action, familiar from
gauge invariant free string field theory \cite{siegel},
\be\label{stueck}
-\int dt d^{p}y
\exp(-\tT/2)\Bigl(\frac{1}{4}\tF_{\hmu\hnu}^{(n)}\tF_{\hmu\hnu}^{(n)}+
\frac{1}{2}n\left(B_\hmu^{(n)}-\p_\hmu\phi^{(n)}\right)^2 \Bigr).
\ee

Using the action in \refb{stueck} we can now study the
reduction of a massive gauge field on the $p$-brane down to a
$(p-1)$-brane. We will show that each massive gauge field
on the $p$-brane gives rise to an equally massive gauge field in
the $(p-1)$-brane, and at equally spaced higher mass levels,
a massive gauge field plus a massive scalar.

  To this end, we consider the $p-1$
brane solution $\tT=w^2-2$ of the $p$ brane field theory.
The massless mode $B_\hmu^{(0)}$ on the $p$-brane works  
exactly like the massless gauge field on the $(p+1)$-brane.  Hence the
fluctuation spectrum for this mode has a massless gauge field on the
$(p-1)$-brane and equally spaced massive vector fields above this.  The
lost massless component, corresponding to the transverse
gauge component,
is compensated
for by the massless tachyon mode, so that the
total number of massless degrees of freedom remains the same.

Next consider the massive vector fields on the $(p-1)$-brane. We let
barred indices $\bar\mu, \bar\nu$,  run over the $p$ world volume
coordinates of
the $(p-1)$-brane.  Let us
impose the axial gauge $B_w^{(n)}=0$ and define new fields
$C_{\bar\mu}^{(n)}=\exp(-w^2/4)B_{\bar\mu}^{(n)}$,
$\tG_{\bar\mu\bar\nu}=\p_{\bar\mu} C_{\bar\nu}-\p_{\bar\nu} C_{\bar\mu}$ and
$\varphi^{(n)}=\exp(-w^2/4)\phi^{(n)}$.  Then the action in
\refb{stueck} becomes
\begin{eqnarray}
\label{stueck2}
&&-\int dt d^{p-1}y\,dw\,
\Biggl\{\frac{1}{4}\tG_{\bar\mu\bar\nu}^{(n)}\tG_{\bar\mu\bar\nu}^{(n)}
+\frac{1}{2}n\left(C_{\bar\mu}^{(n)}-\p_{\bar\mu}\varphi^{(n)}\right)^2
\nonumber\\
&&\qquad+\frac{1}{2}
C_{\bar\mu}^{(n)}\left(-\p_w^2+\frac{1}{4}w^2-\frac{1}{2}\right)C^{(n)}_{\bar\mu}
+\frac{1}{2}n\varphi^{(n)}
\left(-\p_w^2+\frac{1}{4}w^2-\frac{1}{2}\right)\varphi^{(n)}
\Biggr\},
\end{eqnarray}
with the constraint
\be\label{constraint}
\p_w\left(\p_{\bar\mu} B^{(n)}_{\bar\mu} -n\phi^{(n)}\right)=0.
\ee
This then is a nontrivial constraint on all modes that are not in the
ground state of the $w$ coordinate harmonic oscillator.
For the zero modes, as before,
the solution for $C^{(n)}$ and $\varphi^{(n)}$ are such
that the corresponding $B^{(n)}$ and $\phi^{(n)}$ fields
are $w$ independent.

There is still a residual gauge invariance that acts on the lowest
mode of the $C_{\bar\mu}^{(n)}$ fields, hence the lowest mass modes are
massive vector fields on the $p-1$ brane.
The gauge invariance does
not extend to the higher modes.
However, using the constraint in \refb{constraint}
and the equations of motion derived from \refb{stueck2},
one can easily show that the higher modes
satisfy the equations
\begin{eqnarray}\label{Cphieom}
-\p^{\bar\mu} C^{(n,k)}_{\bar\mu}+n\varphi^{(n,k)}&=&0\cr
- \partial_{\bar\mu} \partial^{\bar\mu} C^{(n,k)}_{\bar\nu}
+ (n+k) C^{(n,k)}_{\bar\nu} &=&0\cr
- \partial_{\bar\mu} \partial^{\bar\mu} \varphi^{(n,k)}
+ (n+k) \varphi^{(n,k)} &=&0
\end{eqnarray}
where $k$ is the mass squared
arising from the wavefunctions along the $y$ direction.
Finally, we can define a new field $\widehat C^{(n,k)}_{\bar\mu}$ that is
given by
\be
\widehat C^{(n,k)}_{\bar\mu}=C^{(n,k)}_{\bar\mu}-
\frac{n}{n+k}\p_{\bar\mu}\varphi^{(n,k)}.
\ee
Hence, the first and last
equations in \refb{Cphieom} give
$\p_{\bar\mu}\widehat C^{(n,k)}_{\bar\mu}=0$.
Thus, the higher modes on the $p-1$ brane have  massive vectors
$\widehat C^{(n,k)}_{\bar\mu}$, {\it and}
massive scalars $\varphi^{(n,k)}$.

\sectiono{Comparison to the bosonic D-brane spectrum}

In a field theory with just a tachyon, the fluctuation modes
on the unstable
lump are nondegenerate.  We know of course that open string
modes on D branes
are highly degenerate.  
The original
tachyon model, with the nice feature of having a
discrete spectrum
with the correct spacing between levels,
could not fail to miss most of the states that one would
expect from string theory.  By adding in a gauge field, however, 
there are
more fluctuation modes, enough in fact to start making reasonable
conjectures as to how they
match with the open strings on a D-brane.

The purpose of this section is to do just that and
compare the results for the tachyon and
gauge field fluctuations of
the previous section to results for the D brane spectrum in
bosonic string theory.  We find that our results fit  nicely with a
subclass
of oscillator states for the D brane fluctuations of the bosonic string.
This attests to the power of these models  and
strongly suggests that the original model
can be made into a complete string model by adding the rest of the fields.

In comparing the string theory construction of the spectrum on a 
D25-brane and on a D24-brane, we know that (apart from momentum
modes which differ) the number of physical degrees of freedom are the same
and just rearrange themselves into different representations of the Lorentz
group.  For example the first three levels of the D25 contain the
tachyon, a
massless vector, and a massive symmetric rank two tensor
respectively. On the D24 brane, these turn into a tachyon, a massless vector
and a massless scalar, and at the next level, a 
massive symmetric rank two tensor, a massive vector
and a massive scalar. In our discussion, where the D24 appears
as a solution of the D25 field theory, we have seen that the D25
tachyon localizes to a D24 tachyon, plus a tower of higher
mass squared D24 scalars. The D25 gauge field localizes to a D24 massless
gauge field and a tower of D24 massive gauge fields. Although
we have not made a localization analysis for the massive 
symmetric rank two tensor on the D25, its Stueckelberg formulation
\cite{siegel} suggests strongly that it localizes on the
D24 to an equal mass symmetric rank two tensor (plus additional
states at higher masses).  We therefore see a nice pattern 
appearing, the fluctuations of the tachyon field on the D24 provide
the scalars needed both at the massless and higher mass levels.
In addition, the fluctuations of the gauge field on the D24 provides
the massive gauge field needed at the next level. In summary, modes
``lost" by localization are obtained as fluctuation modes of lower
mass fields. 

\medskip 
In order to provide further insight into this subject we
attempt to identify the particular oscillator states
that correspond to the various localized fields.
To this end consider then the D25 brane of bosonic string theory.
The tachyon state is the
string ground state $|\Omega\rangle=c_{1}|0\rangle$ and
the gauge states are
$\al^\mu_{-1}|\Omega\rangle$.  Notice that under the
twist operation
$\sigma\to\pi-\sigma$, the oscillators transform as
\be\label{twist1}
\al^\mu_{n}\to(-1)^n\al^\mu_n.
\ee
The tachyon state  twist even  and  hence
the gauge field states are twist odd.  In the string
field theory
the fluctuation modes of a field should have the same twist
as the
field itself.  If we now consider lower dimensional D branes,
then the
oscillators with components transverse to the brane have their
twist flipped, so that
\be\label{twist2}
\ta^I_n\to -(-1)^n\ta^I_n,
\ee
 where $\ta^I_n$ refers
to a transverse oscillator.  Hence for the D24 brane, the states
\be\label{states1}
\left(\ta^{25}_{-1}\right)^k|\Omega\rangle
\ee
are all twist even states.  It is
these states that we propose should be identified with
the fluctuations of the tachyon.  For
higher codimension branes,
 the tachyon fluctuations are identified with the
states built with
$\ta^I_{-1}$ oscillators acting on $|\Omega\rangle$. Indeed, with 
$c$ transverse dimensions, this
matches correctly with the spectrum of the $c$-dimensional 
simple harmonic oscillator, which is the spectrum of the tachyon
fluctuations around the codimension $c$ lump solution
(see \cite{0008231} section 5). 

\medskip
Next consider matching the gauge fields on the D24 brane.  We 
suggest the identification 
\be\label{states2}
\al^\hmu_{-1}|\Omega\rangle \,\,\Longleftrightarrow \,\,B_\hmu^{(0)}\,,
\qquad 
\hmu = 0, 1, \cdots 24, 
\ee
with the massless gauge field $B_\hmu^{(0)}$ of the previous section.
For the massive gauge fields we propose
\be\label{states4}
\Bigl\{\,\, \al_{-1}^\mu\left(\ta_{-1}^{25}\right)^{n}|\Omega\rangle\,,\,
\ta_{-2}^{25}\left(\ta_{-1}^{25}\right)^{n-1}|\Omega\rangle\,\, \Bigr\}
\,\,\Longleftrightarrow \,  \Bigl\{ B_\hmu^{(n)}\,, \,
\phi^{(n)} \Bigr\} \quad n= 1, 2, \cdots,   
\ee
where the identification uses the Stueckelberg formulation 
of \refb{stueck}. Note that all the above states are twist odd,
as they ought to be since they arise from the twist odd gauge
field. 

Since we obtained in the previous section the localization of
a massive gauge field (giving an equal mass gauge field and
at each higher level a  gauge field plus a scalar) we can 
examine the gauge fields on the D23, as obtained by localizing
all the gauge fields on the D24.   Again the massless vector
on the D23 is identified 
$\al_{-1}^{\bar \mu}|\Omega\rangle$, with $\bar \mu = 0, 1, \cdots
23$. This is the lowest mode arising from the localization of
the D24 field $B_{\hat\mu}^{(0)}$. The higher massive gauge field modes
arising from the D24 field $B_{\hat\mu}^{(0)}$ are identified as 
\be\label{states5}
\Bigl\{\,\,
\al_{-1}^{\bar\mu}\left(\ta_{-1}^{24}\right)^{k}|\Omega\rangle\,
, \, \ta_{-2}^{24}\left(\ta_{-1}^{24}\right)^{k-1}|\Omega\rangle
\, \Bigr\}
\,\,\Longleftrightarrow \,  \Bigl\{ C_{\bar\mu}^{(0,k)}\,, \,
\varphi^{(0, k)} \Bigr\} \,,  \quad k= 1, 2, \cdots,  
\ee
where comparing with \refb{stueck2} we see that in this
case the $\varphi^{(0, k)}$ are  Stueckelberg fields (the last term
in \refb{stueck2} vanishes and the constraint in \refb{constraint}
is empty).  The remaining gauge field states can be associated with
states arising from the localization of the massive D24 gauge fields
$B_{\hat\mu}^{(n)}$, with $n=1, 2, \cdots$.  They are:
\be\label{states6}
\Bigl\{ \Bigl(\, \al_{-1}^{\bar\mu}\left(\ta_{-1}^{25}\right)^{n}
\hskip-2pt\left(\ta_{-1}^{24}\right)^k\,,
\,\,\ta_{-2}^{25}\left(\ta_{-1}^{25}\right)^{n-1}
\hskip-3pt\left(\ta_{-1}^{24}\right)^k\Bigr),
\, \ta_{-2}^{24}\left(\ta_{-1}^{25}\right)^{k}
\hskip-3pt\left(\ta_{-1}^{24}\right)^{n-1}\, \Bigr\}
\,\Leftrightarrow \,  \Bigl\{ C_{\bar\mu}^{(n,k)}\,, 
\varphi^{(n, k)} \Bigr\} \,,  
\ee
where $k=0, 1, 2, \cdots$. The  fields
$( C_{\bar\mu}^{(n,k)}, \,
\varphi^{(n, k)})$  in \refb{stueck2} define for $k>0$ a massive
gauge field formulated without its Stueckelberg partner,
and a massive scalar.  We see in the above equation that,
as usual, oscillator states in string theory provide for
the Stueckelberg partners, in addition to the scalars.
For $k=0$, there is no extra scalar, and indeed we see that in that 
case the last
state in the above left hand side had already been listed in
\refb{states5}
and therefore is not available.  This concludes our discussion
of the identification of localized gauge states with D-brane
open string states.

\sectiono{Fermion fields in the superstring tachyon model}

For the unstable D9 brane of Type IIA string theory, there are 16
massless
fermionic degrees of freedom described by modes
one 10 dimensional Majorana fermion.  For
the
BPS D8 brane there are 8 massless fermionic
degrees of freedom described by one 9 dimensional
Majorana fermion that pairs up with a massless scalar and gauge
field to
form a supermultiplet \cite{9909062}.
We would like to see how this happens in general
and in the
string
field theory models in particular.

We thus consider the following action for a fermion field
\be\label{fermaction}
S=-\int dtd^9x (\KK' (T) )^2
\left[\frac{i}{2}\overline\psi\Gamma^\mu
\bfpar{\!}_\mu\,
\psi +W(T)\overline\psi\psi\right]\,,
\ee
where
\be
W(T)=- \frac{\KK''(T)}{\KK'(T)},
\ee
and compute the fermion spectrum about the kink solution $T=x$.  The
analysis closely follows classic 
 work on the study of fermions on
a soliton background \cite{Jackiw}.
To this end we define the field $\chi=\KK'(T)\psi$, in which case
the
fermion action in the soliton background reduces to
\be\label{fermfl}
S=-\int dtd^8ydx
\left[\,i\overline\chi\Gamma^\mu \p_\mu\chi
+W(x)\overline\chi\chi\right].
\ee
Hence, the prefactor in \refb{fermaction} plays no role in 
determining the
fluctuation spectrum, leaving its precise form ambiguous 
in this analysis. We adopted the specific form in 
\refb{fermaction} in order to have an action with the same
structural form as the tachyon and gauge field action.

We then choose a basis for the $\Gamma$ matrices
\be\label{gammamat}
\Gamma^x=\left(\begin{array}{ccl}
                             0& iI \\
                               iI & 0 \end{array}\right)\qquad\qquad
\Gamma^\mu=\left(\begin{array}{ccl}
                               \gamma^\mu&0\\
                                   0&-\gamma^\mu
\end{array}\right)\qquad
\mu=0...8,
\ee
where $\gamma^\mu$ refers to the 9 dimensional $\Gamma$ matrices.
The fermion spectrum is determined by finding the eigenvalues of the
linear
equation
\be\label{mateq}
\left(\begin{array}{ccl}
                             W(x)& -\frac{d}{dx} \\
                              -\frac{d}{dx} &
W(x)\end{array}\right)
\left(\begin{array}{cl}
                     \chi_1\\
                     \chi_2\end{array}\right)= m\left(\begin{array}{cl}
                     \chi_1\\
                     -\chi_2\end{array}\right),
\ee
where the sign on the rhs of \refb{mateq} arises because of the form of
$\Gamma^\mu$ in \refb{gammamat}.
Hence, we find that the linear combinations $\chi_\pm=\chi_1\pm\chi_2$
satisfy the equations
\be\label{fermeqs}
\left[-\frac{d^2}{dx^2}+W^2(x)\pm W'(x)-m^2\right]
\chi_\pm = 0 \,
.
\ee
Thus, we find that $\chi_-$ and $\chi_+$ satisfy the two different
Schroedinger equations
\begin{eqnarray}\label{fermschr}
\left(-\frac{d^2}{dx^2}+\frac{\KK'''(x)}{\KK'(x)}-m^2\right)
\chi_- &=& 0\,,
\cr
\left(-\frac{d^2}{dx^2}+2\left(\frac{\KK''(x)}{\KK'(x)}\right)^2
-\frac{\KK'''(x)}{\KK'(x)}-m^2\right)
\chi_+ &=& 0\,.
\end{eqnarray}
The Schroedinger equation for the $\chi_-$ modes is the same as for
the tachyon fluctuations (see eq. \refb{tachschr}).
Hence, in this case
the lowest mode is massless.  The
Schroedinger equation for the $\chi_+$ modes is different from the
tachyon fluctuations, the potential differing by $2W'(x)$.  In general,
$W'(x)$ is a positive definite function, so these modes are all massive.
Therefore, since only half the fermion degrees of freedom
have massless modes,
we find that there are 8 massless fermion modes on the kink
background.

Let us now specialize to
the finite $\ell$
models.  Using eq. \refb{kl} we
have that
\be
W_\ell(T)=\ell\tanh(T).
\ee
Therefore, the potential term in the
second line of \refb{fermschr} is
\be
2\left(\frac{\KK''(x)}{\KK'(x)}\right)^2
-\frac{\KK'''(x)}{\KK'(x)}=\ell^2-\ell(\ell-1)\sech(x)=
2\ell-1+U_{\ell-1}(x),
\ee
in other words the mass spectrum for these fermion modes is derived from
the $\ell-1$ model, shifted by a positive integer.

Next consider the IIA  model, where  $\KK'(T)$
is given by \refb{kinkharm}
and so $W(T)=T/2$.   In this case the action in \refb{fermaction}
becomes
\be\label{fermaction2}
S=-{\cal T}\int dtd^9x \exp(-T^2/2)
\left[\frac{i}{2}\overline\psi\,\Gamma^\mu
\bfpar{\!}_\mu\,
\psi +\frac{1}{2} T\overline\psi\psi\right],
\ee
The  Yukawa coupling in \refb{fermaction2} can be justified by
considering a
three
string amplitude.  Moreover, a term of this sort lifts the fermion
mass to infinity as the tachyon rolls to the closed string vacua at
$x=\pm\infty.$   The mode equations now reduce to
\be\label{fermeqs2}
\left[-\frac{d^2}{dx^2}+\frac{1}{4}x^2\pm\frac{1}{2}-m^2\right]
\chi_\pm = 0 \,
.
\ee
Therefore, the $\chi_+$ modes have their levels
shifted by one unit, with
the lowest mode starting at $m^2=1$.

We have shown that the action in \refb{fermaction} leads to 8 massless
fermion
modes, hence the full model has bose-fermi degeneracy at the massless
level.
However, the massive states do not have this degeneracy.  In order to
achieve
this, as well as a full space-time supersymmetry, one will need to
include an
infinite number of fields in the string field theory action.

\bigskip
\noindent
\sectiono{Higher derivative actions in the superstring tachyon model}

The tachyon action that models the unstable D9 brane of IIA in
\refb{fsaction}
is of course only an approximation to the true string field theory.
Higher derivative terms are indeed present in the complete
action.   The authors of \cite{0010108} have
argued that the
action for a kink solution of the form $T=ux$ has the form\footnote{We
have set
$\alpha'=1$.}
\be\label{bsftaction}
S=-\frac{{\cal T} \sqrt{\pi q}\,4^q
\left(\Gamma(q)\right)^2}{\sqrt{2}\,\Gamma(2q)},\quad T= ux, \quad q
\equiv u^2
\,.
\ee
The action of the kink is minimized if $q\to\infty$, thus
shrinking the width of the kink to zero size.
The action in \refb{bsftaction} clearly has higher derivative terms,
although
it is consistent with an action that only has a
single derivative acting on
{\it each} $T$ field.

\subsection{Higher derivative actions}
\medskip
Let us generalize the action in \refb{fsaction} to be of the
form\footnote{The solvability of tachyonic actions with higher 
derivatives was also noticed by Ashoke Sen \cite{senpv}.}
\be\label{genaction}
S=-{\cal T} \int dt\,d^{p+1}x \,
e^{-T^2/2}f(\partial_\mu T \,\partial^\mu T )\,.
\ee
We will assume for normalization purposes, and in order to have
a nonvanishing term with two derivatives leading to a tachyon, that
\be
\label{conditions}
f(0) = 1\,, \quad f'(0) > 0\,.
\ee
Indeed, with these conditions one readily finds that there is
a tachyon around the $T=0$ vacuum, with mass squared:
\be
\label{tachmass}
M_T^2 = -{1\over 2 f'(0)} \,.
\ee
One can easily see that $T=ux$ is a kink solution provided that
\be\label{soln}
2q f'(q)=f(q)\,, \quad q = u^2.
\ee
This equation should be viewed as a constraint on the choice for
a function
$f$ and on
the value $u$ used in  the kink solution. Finally, we can also
compute for this action
the ratio of brane tensions. By the definition
of the action \refb{genaction}, we have
$\TT_p = \TT$, and splitting $d^{p+1}x =d^{p}ydx$ we find
\be
\TT_{p-1} = \TT \int dx e^{-q \,x^2/2}  f(q) = \TT \sqrt{2\pi\over q}
f(q)\,.
\ee
This leads to
\be
\label{theratio}
{\sqrt{2}\over 2\pi } \, {\TT_{p-1}\over \TT_p} = {1\over \sqrt{\pi}}
\,{f(q)\over
\sqrt{q}} \,,
\ee
which
in string theory must take the value of unity.

As a small check, we verify that this more general setup reproduces the
results obtained before.  For example, the
action
 \refb{fsaction} corresponds to $f(q) = 1+ q$, which solves \refb{soln}
for $q=1 \to u= \pm 1$. In this case $M_T^2 = -1/2$ follows from
\refb{tachmass}, and the ratio in \refb{theratio} takes the value
$2/\sqrt{\pi}$, all this in agreement with \cite{0009246}.

\medskip
We can now
investigate the small fluctuations problem
about the solution
of the general action.  To this end, let
$T=ux+\tT$.  Expanding the action in
\refb{genaction} to second order in $\tT$, using \refb{soln},
redefining the fluctuation field as
$\tT=\wh T \,e^{u^2x^2/4}$,  and integrating by parts, one finds
\begin{eqnarray}\label{flucs} 
S=-{\cal T} \int dt\,d^{p}y\,dx
\Biggl(\,e^{-q\, x^2/2}f(q)+{4f(q)\over q}
\biggl\{ {1\over 2} \partial_\mu\wh T\partial^\mu\wh T
\cr \qquad\qquad
+{1\over 2} \Bigl( 1
+4q^2 \, {f''(q)\over f(q)}
\Bigr)\Bigl[(\partial_x\wh T)^2+\frac{1}{4}x^2q^2{\wh T}^2
-\frac{1}{2}q{\wh T}^2\Bigr]\biggr\}\Biggr).
\end{eqnarray}
Thus the spectrum of the small fluctuations about the kink are given by
the
eigenvalues of the one dimensional harmonic oscillator, as can be seen
from
the second line in the above formula. The lowest
mode is massless and the mass squared spacing is given by
\be\label{spacing}
\Delta m^2=q\left(1+4q^2\frac{f''(q)}{f(q)}\right).
\ee

\medskip
At this point the we can summarize the constraints that can be imposed
on the model presented in eqn.~\refb{genaction}:

\begin{itemize}
\item Getting the string theory tachyon mass (see \refb{tachmass}),

\item Getting the string theory ratio of brane tensions (see
\refb{theratio}),

\item Getting the string theory mass squared spacing (see
\refb{spacing}).

\end{itemize}
In fact, these conditions are not enough to fix uniquely the function
$f$.
On the other hand, it seems clear that the class of models we are
considering
do not lead to the the complete boundary string field theory action for
the tachyon, which is expected to have more general patters of higher
derivative terms. In order to see this we demand that, as in B-SFT,
that the kink solution arises in the limit
$u\to\infty$.   Examining  \refb{soln}
we see that $f(q)\sim \sqrt{q}$ as $q\to\infty$. Furthermore, we see
from
\refb{flucs} that this asymptotic behavior for $f(q)$ is necessary to
insure
a finite mass spectrum.   

We now consider two examples for the function
$f(q)$.  The first is that derived from the action in \refb{bsftaction}.
In this case $f(q)$ is given by
\be
f(q)=\frac{ q\, 4^q \left(\Gamma(q)\right)^2}{2\Gamma(2q)}
=\sqrt{\pi q}+\frac{1}{8}
\frac{\sqrt{\pi}}{\sqrt{q}}+{\rm O}(q^{-3/2}).
\ee
This leads to a mass splitting that is half the expected value.
The likely conclusion is that there are higher derivative terms for the
fluctuations.\footnote{In the complete B-SFT the descent relations
hold exactly \cite{0009103, 0010108,0011002}.}

\medskip
The second example is derived from the modified Born-Infeld action proposed
in \cite{0003122,0003221}.  The tachyon kinetic term is incorporated
into the
action to give the following:
\be\label{tachBI}
S=-\TT\int dt d^{p+1}x\,\, V(T)\sqrt{-\det(\eta_{\mu\nu}
+ F_{\mu\nu}+2\p_\mu T\p_\nu T)}\,.
\ee
In order to match to the potential of B-SFT, we set
\be\label{potrel}
V(T)=\exp(-T^2/2).
\ee
With the gauge field absent, the determinant satisfies
 the relation
\be
-\det(\eta_{\mu\nu}
+ 2\p_\mu T\p_\nu T)=1+2\p_\mu T \p^\mu T.
\ee
Therefore, we can use the above arguments to find the
spectrum for the tachyon field fluctuations, with 
\be\label{fun1}
f(q)=(1+2q)^{1/2}.
\ee
  
Strictly speaking, there is no $q$ that is a 
solution to eq.\refb{soln} for this $f(q)$.  Instead, we can modify the
function to
\be\label{freg}
f_\epsilon(q)=(1+2q)^{{1\over 2}+\epsilon}\,,
\ee
and take the limit $\epsilon\to0$. In this limit, the solution of \refb{soln}
is $q=2/\epsilon\to\infty$, 
and so the width of the brane is shrinking to zero size. 
 Choosing the function in \refb{freg} and taking the limit,
we see that the open string tachyon mass is $m^2=-1/2$.  We also see
from \refb{spacing} that the spacing between mass levels is twice the
magnitude
of the tachyon mass squared.  However, the tension of
the 8-brane as compared to that of the unstable 9-brane is
$2\sqrt{\pi}$ which
is less than the actual value of $\sqrt{2}\pi$.

\subsection{Gauge fluctuations for the modified Born-Infeld action}

In this section we compute the spectrum for the gauge fluctuations on
a kink background for the action in \refb{tachBI}.  
Setting $T$ to the kink solution $T=ux$ and 
expanding the determinant in \refb{tachBI} to second order in $F_{\mu\nu}$,
we have
\be\label{detexp}
-\det(\eta_{\mu\nu}
+ F_{\mu\nu}+2\p_\mu T\p_\nu T)=
(1+2u^2)\left(1-\frac{1}{2}F_{\mu\nu}N^{\nu\lambda}
F_{\lambda\delta}N^{\delta\mu}\right)
+\,\,...\,,
\ee
where 
\be
N^{\mu\nu}=\eta_{\mu\nu}- {2u^2\over 1+ 2u^2} \, \delta_{\mu x}\delta_{\nu
x}\, .
\ee  
Hence, up to second order in fluctuations we find
\be\label{detrel}
-\det(\eta_{\mu\nu}
+ F_{\mu\nu}+2\p_\mu T\p_\nu T)=
1+\frac{1}{2}(1+2u^2)F_{\hmu\hnu}F_{\hmu\hnu}+
F_{\hmu x}F_{\hmu x}\, .
\ee

To find the gauge fluctuations about the kink background, 
we plug \refb{detrel} into \refb{tachBI} and expand to second order,
giving
\be
-\TT\int dt d^{p}ydx\,\, \exp(-u^2x^2/2)(1+2u^2)^{1/2}
\left(\frac{1}{4}F_{\hmu\hnu}F_{\hmu\hnu}+\frac{1}{2(1+2u^2)}
F_{\hmu x}F_{\hmu x}\right).
\ee
Following the arguments  in section 3, we find that the spectrum has
equally spaced levels with spacing
\be
\Delta m^2=\frac{u^2}{1+2u^2}=\frac{q}{1+2q}.
\ee
Therefore, in the limit $q\to\infty$, the spacing of the levels is half 
 the expected value.  

While not the desired value, we note that the spacing we found is finite.
If the Born-Infeld action was of the usual type, without the tachyon kinetic
term, then for a kink of the form $T=ux$, the spacing would have been
$u^2=q$.  If $q\to\infty$, as is expected in B-SFT, this would 
push the spacing to infinity.   That it stays finite seems to provide
some  
support for the form of the action in \refb{tachBI}.

\subsection{Higher codimension branes}

We can also consider higher codimension branes. In the models for
the bosonic string tachyon such solutions existed exploiting the
fact that the tachyon potential is unbounded below.
Since the tachyon potential for the
superstring is positive definite we need a different way to avoid
Derrick's no-go result for higher codimension
branes. This would seem to require several tachyon fields and gauge
fields.
However, Derrick's theorem assumes a standard kinetic operator for the
scalar fields.
If there are higher derivative terms then higher codimension objects
appear
to
be possible when we have several tachyon fields, even without exciting
the
gauge fields.

In order to construct these higher codimension branes, we thus
start with more than one D9 brane.  If we have $N$ such branes,
then the tachyons  transform in the adjoint of $U(N)$.   For a
codimension $d$ brane, the tachyon configuration has the
form \cite{senconj,9810188,9812135,0010108}
\be\label{tachyond}
T=\Gamma_m \, u\, {x^m}\,,
\ee
where the index $m$ refers to the transverse directions and the
$\Gamma_m$ are the $d$ dimensional Dirac matrices.  The tachyon
configuration in \refb{tachyond} is a solution to the equations of
motion for the action
\be\label{daction}
S = -{\cal T} \int dt\,d^{p+1}x \,
{\rm Tr}\left[e^{-T^2/2} f(\partial_\mu T \,\partial^\mu T )\right]\,.
\ee
with $f$ satisfying
\be\label{fdeq}
2\frac{q}{d}f'(q)=f(q),
\ee
where $q=du^2$.  If we use the two derivative action in
\refb{fsaction}, then we see that there is no solution for $d>2$ and
that in the limit $d\to2$, $q$ is pushed out to infinity.

However, including higher derivative terms leads to solutions with
finite values of $q$.  For instance, suppose that we choose $f(q)$ to
be
\be\label{higherderiv}
f(q)=\left(1+q/d\right)^d.
\ee
This choice for $f(q)$ leads to  a tachyon mass with $m^2=-1/2$.
It has a solution at $q=d$, which corresponds to $u=1$.
Computing the tension of the codimension $d$ brane, we find that
\be\label{dtension}
T_{9-d}=2^{d/2}2^d(2\pi)^{d/2}{\cal T}.
\ee
The total tension of the 9 branes is $2^{d/2}{\cal T}$, which
takes into account the multiple unstable branes or brane-antibrane
pairs, depending on whether $d$ is odd or even.  Hence the ratio of
tensions to a single IIB 9 brane is
\be\label{ratio}
 \frac{T_{9-d}}{T_9}=2^d(4\pi)^{d/2}=
\left(\frac{4}{\pi}\right)^{d/2}(2\pi)^d.
\ee
Hence, this choice for $f(q)$ leads to constant descent relations and
can basically be thought of as the extension to the two derivative
truncation for lower dimensional branes.
\bigskip

\section{Conclusions}

We have extended the previously constructed
superstring tachyon effective
theory \cite{0009246} to include gauge fields and fermions. The
complete model, assembled from \refb{finmodg} and \refb{fermaction}
reads
\ben
\label{completess}
S = - \int \hskip-3pt dt d^{p+1}x \Bigl({\cal K}' ( T)\Bigr)^2 \,
\Bigl[\,  (\partial T)^2  +1 \, + {1\over 4}
F^{\mu\nu}F_{\mu\nu} +\frac{i}{2}\overline\psi\,\Gamma^\mu
\bfpar{\!}_\mu\,
\psi +W(T)\,\overline\psi\psi \Bigr]\,,
\een
where  $\KK' (T) = \sqrt{\TT} \exp (-T^2/4)$,
$W(T) = T/2$, and we work with $p=8$ for the unstable D9 of IIA.
This action gives the correct tachyon mass, and correct mass
spectrum for the tachyon, gauge field and fermion field fluctuations
around the kink solution representing a codimension one brane. 
On the other hand, the descent relation for the tension is not exactly
that of string theory.  As we
have explained, the above 
interactions for gauge fields and fermion fields
result in solvable spectra around brane configurations, 
but any
action for gauge fields or fermions that is equivalent to this action, 
up to quadratic order in
those fields,  will  have the
 same solvability properties and exactly  the same spectrum. This
includes, in particular,  Born-Infeld actions of the form
\be
S= -\int \hskip-3pt dt d^{p+1}x\Bigl({\cal K}' (
T)\Bigr)^2\sqrt{-\hbox{det}
(\eta_{\mu\nu}+ F_{\mu\nu}) } + \cdots \,,
\ee
where the dots denote tachyon and fermion kinetic terms. In fact, our
analysis 
of actions with higher derivatives in  section 6  shows that
solvability is
still possible if, following \cite{0003122,0003221},  we include the
tachyon kinetic term inside the square root
giving
\be
S= -\int \hskip-3pt dt d^{p+1}x\Bigl({\cal K}' (
T)\Bigr)^2\sqrt{-\hbox{det}
(\eta_{\mu\nu}+ F_{\mu\nu} +
2\,\p_\mu T \p_\nu T )} + \, \hbox{fermions}  \,. 
\ee
Details of the solutions and spectra, however, differ from that
of \refb{completess}, since the tachyon background is changed 
(it is $T=ux$, with $u\to \infty$).  In particular, the level
spacing for the gauge field fluctuations is half the expected one.

\medskip
For the bosonic string theory tachyon model the
two alternative forms of the action are based on equations
\refb{bgaction} and \refb{hjk}
\be\label{finbosact}
S = - \int dt d^{p+1}x ( \LL'(T))^2 \,\,\Bigl( \,
{1\over 2} (\partial T)^2  + 2 T \,  +
\Biggl\{ {2T  \atop ~1} \Biggr\}{1\over 4} F_{\mu\nu}
F^{\mu\nu}\Bigr) \,,
\ee
with $\LL'(T)=\frac{1}{2}\sqrt{{\cal T}e}\exp(-T/4)$.
The first form is  Born-Infeld compatible and has finite
$\ell$ solvable counterparts. The spectrum
of gauge fluctuations, however, has a mass spacing which
is twice the expected one. The second form is not Born-Infeld
compatible and does not
have finite $\ell$ solvable counterparts. On the other hand
the mass spacing is the expected one. Note that neither form
is compatible with tachyon kinetic terms inside the Born-Infeld
square root, since the prefactor in the tachyon kinetic term
does not coincide with the tachyon potential.
It is not completely clear how to decide between these two
possibilities, but it seems reasonable to choose the form
with correct mass spacing. As reviewed in the introduction, both
forms have appeared in B-SFT studies. 

\medskip
In section (6.2) we saw that one could  construct stable solutions
on the worldvolume of coincident branes
representing branes with codimension greater than one {\it without}
turning on 
gauge fields.  This was possible because the tachyon field theory model
used had higher
derivative  terms for the tachyon, and is why such solutions could be found
also in B-SFT \cite{0010108}.  The existence of such solutions,
however, is somewhat puzzling as earlier discussions of 
configurations on the world volume of coincident D-branes 
leading to a lower dimensional stable branes appear to require
non-trivial gauge field backgrounds
\cite{senconj,9810188,9812135}.

\medskip  
The surprising and pleasant features of the original tachyon models
\cite{0008231,0009246} were that simple field theory interactions
exactly described a situation where the tachyon could dissappear 
completely in the stable vacuum and where nontrivial configurations
representing branes existed with a fluctuation spectrum of stringy type.
In retrospect, we see that such properties essentially guaranteed 
agreement between these 
models and the two-derivative truncations of B-SFT
formulations of tachyon dynamics.  We have seen in this paper that these
pleasant features nicely extend to models including the interactions of the
tachyon with gauge and fermion fields. 

\bigskip\bigskip
\noindent {\bf Acknowledgments}:
J.A.M. would like to thank the CTP at MIT and the theory group at
Harvard for hospitality during the course of this work.  We are
grateful to
Jeffrey Goldstone for discussions on fermion localization
on kinks. We also wish to thank Igor
Klebanov and  Leonardo Rastelli for their comments on
gauge invariance in B-SFT. Finally, we would  like to thank
Ashoke Sen for many useful remarks and discussions. This work was
supported in part by DOE contract
\#DE-FC02-94ER40818.

\end{document}